# Mathematical Basis for Analyzing Superconducting Phase Transitions Using Catastrophe Theory

Jiu Hui Wu [1*], Hua Tian [1], and Kejiang Zhou [2]

[1]*School of Mechanical Engineering, Xi'an Jiaotong University,*

*& State Key Laboratory for Strength and Vibration of Mechanical Structures, Xi'an 710049, China*

[2] *Huzhou Institute of Zhejiang University, Huzhou 313000, China*

* *ejhwu@xjtu.edu.cn*

**Abstract**: We establish a rigorous mathematical bridge from quantum many-body path integrals to the cusp catastrophe model by Lyapunov-Schmidt reduction, which provides a theoretical foundation for analyzing superconducting phase transition using the catastrophe theory. First, it is proved that, near the critical point the infinite-dimensional effective action is diffeomorphic to a finite-dimensional catastrophe. Secondly, starting from Ginzburg-Landau free energy functional, the Euler-Lagrange partial differential equation can be reduced to the cusp catastrophe model. Thirdly, the fermionic imaginary-time path integral to the cusp catastrophe is derived through the Hubbard-Stratonovich transformation, Matsubara frequency expansion, and Grassmann algebra. Furthermore, we connect this framework with the adsorption potential theory we proposed, elucidating the catastrophic topological nature of the electron pairing mechanism in high-temperature superconductivity. The precise microscopic derivation of the adsorption potential from first-principles electronic structure calculations would strengthen the predictive power of the theory.

## 1. Introduction

The phase transition of quantum many-body systems is a core topic in condensed matter physics. From the Bardeen-Cooper-Schrieffer (BCS) theory of super-conductivity [1] to the Ginzburg-Landau (GL) phenomenological theory [2], traditional methods have achieved tremendous success in describing phenomena such as low-temperature superconductivity and superfluidity. However, with the discovery of strongly correlated systems such as high-temperature superconducting cuprates [3], iron-based superconductors [4], and nickel-based superconductors [5], the BCS-GL framework based on weak-coupling perturbation theory faces fundamental challenges: the transition temperatures of these systems far exceed the McMillan limit (~40 K) [6], and their pairing mechanisms cannot be explained solely by electron-phonon interactions.

In recent years, in-depth experimental and theoretical studies have revealed several universal features of high-temperature superconductors: strong electron correlations [7], d-wave pairing symmetry [8], pseudogap behavior [9], and non-Fermi liquid normal state [10]. These "anomalous" phenomena have prompted the academic community to seek new theoretical paradigms beyond the BCS framework. Meanwhile, the emergence of concepts such as quantum criticality [11], topological order [12], and many-body localization [13] has further highlighted the urgent need to develop non-perturbative, universal theories of phase transitions.

Finite-temperature quantum field theory provides a systematic framework for dealing with interacting many-body systems. Based on the Matsubara imaginary-time path integral [14], the partition function has been calculated out. For fermion systems, the Hubbard-Stratonovich transformation [15] decouples the four-fermion interaction by introducing an auxiliary field, and then performs Gaussian integration over the fermionic degrees of freedom to obtain the effective action [16]. This procedure has been widely applied in systems such as BCS superconductors [1], charge density waves [17], and spin density waves [18].

However, the effective action is still an infinite-dimensional functional, and its direct solution or qualitative analysis faces fundamental difficulties. Although the standard Ginzburg-Landau expansion [2] can derive a low-energy effective theory, the rationality of its truncation scheme and the systematic treatment of higher-order fluctuations lack a rigorous mathematical foundation. In particular, when the system approaches the critical point, the divergence of critical fluctuations causes the perturbative expansion to fail, urgently requiring non-perturbative methods.

Lyapunov-Schmidt (L-S) reduction [19,20] provides a rigorous mathematical tool for dealing with infinite-dimensional nonlinear equations. The core idea is that, at the critical point, the kernel of the linearized operator $L$, ker$L$ (the zero mode space), captures the 'dangerous' directions of the system, while the directions in the orthogonal complement space are 'stable' and can be eliminated using the implicit function theorem. If $L$ satisfies the Fredholm properties (finite-dimensional kernel, closed range, finite-dimensional cokernel), then the original infinite-dimensional problem can be rigorously reduced to a finite-dimensional bifurcation equation [21].

For self-adjoint elliptic operators, the elliptic regularity theory [22] ensures that they have a compact parametrix, thus the spectrum is discrete and the zero-mode space is finite-dimensional [23]. Under periodic boundary conditions, the L-S reduction yields a finite-dimensional equation for the order parameter amplitude [24].

Thom's catastrophe theory [25,26], based on the principle of structural stability, provides a universal classification of singularities in gradient systems. Thom's classification theorem asserts that when the number of state variables $n \leq 2$ and the number of control parameters $m \leq 4$, any structurally stable potential function germ is diffeomorphic to one of the seven elementary catastrophes. The mathematical foundation of this classification lies in the finite determinacy theorem and the theory of universal unfolding [27,28].

The application of catastrophe theory in physics began with Zeeman's explanation of phase transitions and bifurcation phenomena [29]. Arnold further linked the classification of singularities of caustic surfaces with the ADE classification of simple Lie algebras [30], revealing the deep connections between catastrophe structures, symplectic geometry, and wavefront propagation. In condensed matter physics, cusp catastrophes have been used to describe liquid-gas phase transitions [31], ferroelectric phase transitions [32], and liquid crystal phase transitions [33].

More recently, the connection between catastrophe theory and quantum mechanics has been revisited. Wu et al., based on the principle of structural stability, applied the cusp catastrophe model to quantum thermodynamic phase transitions, deriving a universal quantum equation of state that includes Bose and Fermi gases [34]. This framework was further extended to superconducting phase transitions, proposing the concept of 'adsorption potential' to explain the pairing mechanism of high-temperature superconductivity [35,36]. However, these works are mainly presented as phenomenological applications, lacking a rigorous derivation chain from quantum many-body first principles to the catastrophe model.

In 2026, Lohmiller and Slotine [37] proposed the exact construction of quantum wave functions based on the classical principle of least action. They demonstrated that the solutions of the Schrödinger equation can be strictly expressed as a superposition of classical extremal paths, with each path weighted by the classical density, without the need for the infinitely many non-classical jagged paths in Feynman path integrals. This result provides a new exact framework for classical-quantum correspondence and implies the inherent nature of cusp structures in quantum mechanics. The deep connection between this work and catastrophe theory lies in the fact that the behavior of the multi-valued branches of the classical action at branch points precisely corresponds to the folding and jumping of equilibrium surfaces in catastrophe theory; meanwhile, the condition of structural stability ensures the topological invariance of quantum observables.

In recent years, research on high-temperature superconductivity has shown a trend of diverse breakthroughs. In the field of nickel-based superconductors, the ambient-pressure superconductivity of infinite-layer nickelates has been confirmed [5, 38], and the coupling of their Higgs mode with the pseudogap provides new clues for the pairing mechanism [39,40]. Universal scaling relations between quantum critical fluctuations and strange metal resistivity have been simultaneously observed in nickelates and cuprates [41], suggesting that the two systems may share a common quantum critical origin. In the case of cuprates, the superconducting transition temperature of $YBa_2Cu_3O_{7-\delta}$ under high pressure has been raised above 100 K [42], while the Z-entropy theory attempts to unify the BCS theory, density functional theory, and strongly correlated systems [43]. Hydride superconductors have achieved near-room-temperature superconductivity (Tc ~ 200-288 K) under megabar pressures [44,45], with their mechanism attributed to strong electron-phonon coupling and pressure-induced nearly free electron behavior [46]. Furthermore, interface superconductivity continues to exhibit novel phenomena in oxide heterostructures [47,48] and two-dimensional crystal films [49], providing a platform for the artificial design of superconducting states.

This paper aims to establish a rigorous mathematical chain from finite-temperature quantum many-body path integrals to cusp catastrophe models. Taking fermionic systems as an example, through the Hubbard-Stratonovich transformation, Matsubara frequency expansion, and Lyapunov-Schmidt reduction, we systematically prove that, near the critical point, the effective action is diffeomorphic to a cusp catastrophe model. Furthermore, we connect this framework with the adsorption potential theory, elucidating the catastrophic topological nature of the electron pairing mechanism in high-temperature superconductivity.

## 2. Mathematical Foundation

In quantum field theory and path integrals, we are dealing with infinite-dimensional problems, i.e. the effective action $\Gamma[\phi]$ is a functional with the variable $\phi$(r) being infinite-dimensional, and the Euler-Lagrange equation $\delta\Gamma/\delta\phi = 0$ is an infinite-dimensional nonlinear equation. Directly solving it or classifying its singularity structure is almost impossible.

### 2.1 Background of the problem

Considering a nonlinear functional equation

$$F(u, \lambda) = 0 \tag{1}$$

where $u \in X$, $X$ is the Banach infinite-dimensional space, $\lambda \in \mathbb{R}^k$ is the $k$-dimensional control parameter, and $F: X \times \mathbb{R}^k \to Y$ is a smooth map.

For the critical point condition $F(u_0, \lambda_0) = 0$, the linearization $L = D_u F(u_0, \lambda_0)$ has a nontrivial kernel, i.e. $\ker L = \{\varphi \in X | L\varphi = 0\}$, spanned by the zero modes $\varphi_1, \cdots, \varphi_n$.

Near the critical point (singularity), the truly 'dangerous' directions are only a finite number of zero-mode directions. The remaining infinite directions are non-degenerate (Morse type) and can be solved away using the implicit function theorem. While retaining all bifurcation, singularity, and solution branch information, Lyapunov-Schmidt reduction will be used to equivalently reduce the original infinite-dimensional equation to an algebraic equation defined on a finite-dimensional space.

### 2.2 Lyapunov-Schmidt reduction

By the elliptic regularity theory, $L$ is a Fredholm operator of index zero, which ensures that the residual kernel dimension matches the kernel dimension. We decompose the spaces as $X = \ker L \oplus M$ and $Y = \mathrm{range} L \oplus N$, where $\ker L$ is the kernel (or null space) of $L$, and $N \cong \mathrm{coker} L = Y/\mathrm{range} L$ is a complement to the range (or image). Thus, near the critical point there is

$$u = u_0 + \sum_{i=1}^{n} x_i \varphi_i + w \tag{2}$$

where $v = \sum_{i=1}^{n} x_i \varphi_i \in \ker L$, in which the zero mode directions correspond to the state variables $\boldsymbol{x} = (x_1, \cdots, x_n) \in \mathbb{R}^n$, and $w \in M = (\ker L)^{\perp}$.

According to the Lyapunov-Schmidt reduction, Eq. (1) can be decomposed into the following two equations

$$\begin{cases} P \cdot F(u_0 + v + w, \lambda) = 0 & \text{(Range equation)} \\ (I - P) \cdot F(u_0 + v + w, \lambda) = 0 & (\mathit{Cokernel}\ \text{equation}) \end{cases} \tag{3}$$

where $P: Y \to \mathrm{range}\, L$ is a projection operator that projects $F(u, \lambda)$ onto the range, $(I - P): Y \to N \cong \mathrm{coker} L$ that projects onto the cokernel, and $I$ is the identity operator on Y. The finite-dimensionality of $\mathrm{coker} L$ ensures that $(I - P)$ projects onto a finite-dimensional space, which is the basis of the Lyapunov-Schmidt reduction.

For the range equation in the non-critical direction, because $L|_M: M \to \text{range}\, L$ is isomorphism (nontrivial), by the implicit function theorem, there exists a unique smooth function $w = W(v, \lambda)$ such that $P \cdot F(u_0 + v + W(v, \lambda), \lambda) = 0$, in which $W(0, \lambda_0) = 0$, and $D_v W(0, \lambda_0) = 0$. Then substituting $w = W(v, \lambda)$ into the cokernel equation, we can obtain the bifurcation equation

$$g(\boldsymbol{x}; \lambda) = (I - P) \cdot F\left(u_0 + \sum_{i=1}^{n} x_i \varphi_i + W\left(\sum_{i=1}^{n} x_i \varphi_i, \lambda\right), \lambda\right) = 0 \quad (4)$$

where $g(\boldsymbol{x}; \lambda)$ is the reduced mapping on the finite-dimensional state variable space $\ker L \cong \mathbb{R}^n$.

### 2.3 Catastrophe models obtained from the Reduced equation

Any phase transition is driven by physical instability. In order to examine whether the structure of phase trajectory motion is stable, a small amount can be added to the system's state equation to perturb the entire vector field. If the phase diagram of the system after perturbation is topologically equivalent to the initial phase diagram in terms of orbit, then the original system is structurally stable.

Due to the Lyapunov-Schmidt reduction, the infinite-dimensional effective action can be reduced to a finite-dimensional model near the critical point. Further, for a gradient system $F(x, \lambda) = -\nabla_x V(\boldsymbol{x}; \lambda)$ in Eq. (1), Thom proposed the catastrophe theory based on the potential function $V(\boldsymbol{x}; \lambda)$ satisfying the equilibrium surface equation $\nabla_x V(\boldsymbol{x}; \lambda) = 0$ and the catastrophe set equation $\det\left|\frac{\partial^2 V(\boldsymbol{x};\lambda)}{\partial x_i x_j}\right| = 0$ (i.e., Hessian matrix degenerate point) in the neighborhood of the zero solution, where the state variable $\boldsymbol{x} \in \mathbb{R}^n$ and the control parameter $\lambda \in \mathbb{R}^k$.

According to Thom's classification theorem, as long as $n \leq 2$ and $k \leq 4$, various catastrophe processes in nature can be grasped using seven basic potential function models. Catastrophe phenomena belonging to the same model have diffeomorphic catastrophe sets.

**Theorem 2.3.1** If $F(u, \lambda)$ satisfies the finite codimension condition, the singular behavior of $g(\boldsymbol{x}; \lambda)$ near $(\boldsymbol{0}, \boldsymbol{\lambda}_0)$ is completely determined by the gradient $\nabla_x V(\boldsymbol{x}; \lambda) = 0$ of some Thom elementary catastrophe model $V(\boldsymbol{x}; \lambda)$, i.e. $g(\boldsymbol{x}; \lambda) \cong \nabla_x V(\boldsymbol{x}; \mathrm{m}(\lambda))$, where $\cong$ denotes contact equivalence, and $\mathrm{m}(\lambda)$ is a reparameterization of the control parameter.

Proof. This theorem can be verified directly through the above process.

Therefore, for any continuous phase transition with control parameters $m \leq 4$, near the critical point, its infinite-dimensional effective action $\Gamma[\phi]$ must reduce to a finite-dimensional catastrophe model $V(\boldsymbol{x}; \mathrm{m})$, exhibiting a bifurcation structure diffeomorphic to one of the seven elementary catastrophes.

## 3. Cusp Catastrophe from Ginzburg-Landau free energy functional

Lyapunov-Schmidt reduction is a mathematical bridge connecting infinite-dimensional quantum field theory and finite-dimensional catastrophe theory. In this section, near the phase transition critical point $T \to T_c$, the Euler-Lagrange partial differential equation can be reduced to a finite-dimensional bifurcation equation through Lyapunov-Schmidt reduction, and then singularity classification can be further performed by Thom catastrophe theory.

In the field of superconductivity, the Ginzburg-Landau free energy functional is

$$\Gamma[\psi(\boldsymbol{r})] = \int_V dV\left[a(\mathrm{T})|\psi|^2 + \frac{b}{2}|\psi|^4 + \frac{\hbar^2}{2m^*}|\nabla\psi|^2\right] \tag{5}$$

and the corresponding Euler-Lagrange equation by variation is

$$-\frac{\hbar^2}{2m^*}\nabla^2\psi + a(\mathrm{T})\psi + b|\psi|^2\psi = 0 \tag{6}$$

where $\psi(\boldsymbol{r})$ is the order parameter (macroscopic wave function of Cooper pairs), $a(T) = a_0(T - T_c)$ is temperature coefficient with $a_0 > 0$, $m^* = 2m_e$ is the effective mass of Cooper pairs, $b > 0$ is the quartic coefficient (ensuring stability).

At the critical point $T = T_c$ $(a(T) = 0)$, the linearizer $L = -\frac{\hbar^2}{2m^*}\nabla^2$. Under periodic boundary conditions, $\psi = \mathrm{const}$, and the zero mode directions of the linearizer span the kernel ker$L$=span$\{1\}$ (the constant mode) yielding a one-dimensional state variable.

According to the Lyapunov-Schmidt reduction, $\psi(\boldsymbol{r}) = x \cdot \varphi_0 + w(\boldsymbol{r})$, where $\int \boldsymbol{w d^n r} = \boldsymbol{0}$ ($w$ is orthogonal to the constant mode, i.e. $w \perp \varphi_0$) and $\varphi_0 = 1/\sqrt{V}$ is a normalization constant. Solving the non-critical direction equation (projection onto the orthogonal complement of ker$L$) for $w$:

$$Pw: \quad -\frac{\hbar^2}{2m^*}\nabla^2 w + a(\mathrm{T})w + bP \cdot |x\varphi_0 + w|^2(x\varphi_0 + w) = 0 \tag{7}$$

Here the operator $P$ is projected onto range $L$, and $a(\mathrm{T})(x\varphi_0) + b(x\varphi_0)^3 \in \ker L$ is constant, there is $P \cdot [a(\mathrm{T})(x\varphi_0) + b(x\varphi_0)^3] = 0$. By the implicit function theorem, from Eq.(7) we can obtain that $w = W(x, T)$ with $W(0, T_c) = 0$, and $W = O(x^3)$ at the lowest level.

Further substitute $w = W(x, T)$ into the cokernel equation (projected onto the constant mode). The projection ($I$–$P$) becomes an integral, essentially extracting the component in the zero-mode direction (constant function). Physically, this is the spatial averaging or coarse-graining, reducing the microscopic field theory to macroscopic order parameter equations. Thus there is

$$g(x; \lambda) = \frac{1}{V}\int [a(\mathrm{T})(x\varphi_0 + W) + b|x\varphi_0 + W|^2(x\varphi_0 + W)]\, \boldsymbol{d^n r} = 0 \tag{8a}$$

Expanding Eq.(8) to the lowest order ($O(x^3)$ can be ignored), we have

$$g(x; \lambda) = bx^3 + a^{'}(\mathrm{T})x + c = 0 \tag{8b}$$

where $a^{'} = a(\mathrm{T})/V^{3/2}$ and $c = c(T)$.

Equation (8b) is precisely the equilibrium surface equation of the cusp catastrophe model.

From the above process, the Lyapunov-Schmidt reduction ensures that: regardless of how many control parameters the original system has (such as chemical potential, stress, doping, etc.), near the critical point, if the truly relevant directions do not exceed 4, its phase transition process must fall into the Thom classification.

# 4. Cusp Catastrophe from Quantum many-body system

## 4.1 Construction of the Fermion Imaginary-Time Action

For interacting fermion systems, the finite-temperature partition function is expressed using imaginary-time path integrals as

$$Z = \int D[\bar{c}, c] e^{-S_E[\bar{c},c]/\hbar}, \tag{9}$$

where $c_\sigma(\boldsymbol{r}, \tau)$ and $\bar{c}_\sigma(\boldsymbol{r}, \tau)$ are both Grassmann variables, satisfying the opposition relation $\{c_\sigma(\boldsymbol{r}, \tau), c_{\sigma'}(\boldsymbol{r}', \tau')\} = 0$, and the Euclidean action is

$$S_E = \int_0^{\beta\hbar} d\tau \int d^n r \left[\bar{c}_\sigma \left(\hbar\partial_\tau - \frac{\hbar^2\nabla^2}{2m^*} - \mu\right) c_\sigma + U\bar{c}_\uparrow\bar{c}_\downarrow c_\downarrow c_\uparrow\right] \tag{10}$$

where $\beta = 1/(k_B T)$, $\mu$ is the chemical potential, $m^*$ is the effective mass, $\partial_\tau = \partial/\partial\tau$ is imaginary time derivative, $U$ is in-orbital Coulomb repulsion energy, $c$ and $\bar{c}$ are independent Grassmann variables (non-complex conjugate) satisfying the anti-periodic imaginary-time boundary $c(\tau + \beta\hbar) =- c(\tau)$.

## 4.2 Hubbard-Stratonovich transformation

Introducing the complex Hubbard-Stratonovich auxiliary field, there is

$$e^{-U\bar{c}_\uparrow\bar{c}_\downarrow c_\downarrow c_\uparrow} = \int D[\Delta^*, \Delta] e^{-\left[\frac{|\Delta|^2}{U} + \Delta^* c_\downarrow c_\uparrow + \Delta\bar{c}_\uparrow\bar{c}_\downarrow\right]} \tag{11}$$

where $\Delta \sim \langle c_\downarrow c_\uparrow \rangle$ means that the superconducting energy gap field is proportional to the Cooper pair condensation amplitude. Thus the action after Hubbard-Stratonovich transformation is

$$S_E[\bar{c}, c, \Delta^*, \Delta] = \int_0^{\beta\hbar} d\tau \int d^n r \left[\frac{|\Delta|^2}{U} + \bar{c}_\sigma {G_0}^{-1} c_\sigma + \Delta^* c_\downarrow c_\uparrow + \Delta\bar{c}_\uparrow\bar{c}_\downarrow\right] \tag{12}$$

where ${G_0}^{-1} = \hbar\partial_\tau - \frac{\hbar^2\nabla^2}{2m^*} - \mu$ is the free fermion inverse propagator.

## 4.3 Integrating out fermions and obtaining the effective action

Because the action is bilinear in fermions, the integral over $c$ and $\bar{c}$ is strictly a Gaussian integral, and there is

$$\int D[\bar{c}, c] e^{-\bar{c}\mathcal{G}^{-1}c} = \det(\mathcal{G}^{-1}) = e^{\mathrm{Tr}(\ln \mathcal{G}^{-1})} \tag{13}$$

where $\mathcal{G}^{-1} = \begin{pmatrix} {G_0}^{-1} & \Delta \\ \Delta^* & -{G_0}^{-1} \end{pmatrix}$ is the Nambu-Gorkov matrix in the Nambu spinor basis $\psi = (c_\uparrow, \bar{c}_\downarrow)^T$.

Thus the effective action only including $\Delta$ is

$$S_{eff}[\Delta^*,\Delta] = \int d\tau d^n r \frac{|\Delta|^2}{U} - \hbar \mathrm{Tr}(\ln \mathcal{G}^{-1}). \tag{14a}$$

Since there are

$$\mathrm{Tr}(\ln \mathcal{G}^{-1}) = \mathrm{Tr}\left(\ln {G_0}^{-1}\right) + \mathrm{Tr}[\ln(1 + G_0 \Delta G_0 \Delta^*)]$$

$$= \mathrm{const} + \sum_{\mathrm{n}=1}^{\infty} \frac{(-1)^{\mathrm{n}+1}}{\mathrm{n}} \mathrm{Tr}[(G_0 \Delta G_0 \Delta^*)^{\mathrm{n}}] \tag{14b}$$

in which

$$\mathrm{Tr}[G_0 \Delta G_0 \Delta^*] = \sum_{\omega_\mathrm{n},k} G_0(\omega_\mathrm{n},k)\, G_0(-\omega_\mathrm{n},-k) \left|\Delta_{\omega_\mathrm{n},k}\right|^2 \propto \left(\frac{1}{U_c} - \frac{1}{U}\right) |\Delta|^2 = a(T)|\Delta|^2$$

where $U_c$ is the critical interaction strength and $\omega_\mathrm{n} = \frac{(2\mathrm{n}+1)\pi}{\beta\hbar}$ is Matsubara frequency, and

$$\mathrm{Tr}[(G_0 \Delta G_0 \Delta^*)^2] \propto b(\mathrm{T})|\Delta|^4,$$

we have

$$S_{eff}[\Delta] = \beta\hbar \int d^n r \left[a(T)|\Delta|^2 + \frac{b(\mathrm{T})}{2}|\Delta|^4 + \frac{\hbar^2}{2m^*}|\nabla\Delta|^2 + \cdots\right] \tag{14c}$$

In addition, by Matsubara expression there is

$$\Delta(\boldsymbol{r},\tau) = \Delta_0(\boldsymbol{r}) + \sum_{n\neq 0} e^{-i\omega_n \tau} \Delta_{\omega_n}(\boldsymbol{r}) \tag{14d}$$

Because $\Delta$ is a bosonic field (composed of two fermions), the imaginary-time boundary of $\Delta$ is periodic satisfying $\Delta(\tau + \beta\hbar) = \Delta(\tau)$. Thus $\Delta$ can have a zero-frequency mode $\omega_n = 0$.

Therefore, for the Gaussian integral over $\Delta_{\omega_n \neq 0}$, the static effective action is obtained as

$$S_{eff}[{\Delta_0}^*, \Delta_0] = \beta\hbar \int d^n r \left[a(T)|\Delta_0|^2 + \frac{b(\mathrm{T})}{2}|\Delta_0|^4 + \frac{\hbar^2}{2m^*}|\nabla\Delta_0|^2\right] \tag{14e}$$

**4.4 Lyapunov-Schmidt reduction in Fermion case**

The Euler-Lagrange equation of Eq.(14e) is

$$\frac{\delta S_{eff}}{\delta \Delta_0^*} = a(T)\Delta_0 + b(\mathrm{T})|\Delta_0|^2 \Delta_0 - \frac{\hbar^2}{2m^*}\nabla^2 \Delta_0 = 0 \tag{15}$$

At $\Delta_0 = 0$, that is near the critical point $T = Tc$ with $a(T) = 0$, the linearizer $L = a(T) - \frac{\hbar^2}{2m^*}\nabla^2 = -\frac{\hbar^2}{2m^*}\nabla^2$. According to the zero-mode condition $L\varphi = 0$ under periodic boundary conditions, there is a zero-mode $\varphi$ = const (i.e., the wavenumber $k$ = 0), and thus ker$L$=span{1}.

Let $\Delta_0(\boldsymbol{r}) = x \cdot \varphi_0 + W(\boldsymbol{r})$, where $x$ is a complex amplitude, $W(\boldsymbol{r})$ contains the modulus where $k \neq 0$ and $W \perp \varphi_0$. Thus by the range equation (projected to $k \neq 0$) $P \cdot \frac{\delta S_{eff}}{\delta \Delta^*} = 0$, we can obtain $W = W(x,T)$. Further from the cokernel

equation

$$g(x;T) = (I-P)\cdot\frac{\delta S_{eff}}{\delta\Delta^*} = \frac{1}{V}\int\frac{\delta S_{eff}}{\delta\Delta^*}d^n r = 0 \tag{16}$$

there is $g(x;T) = a(T)x + b(T)|x|^2x + O(x^3) = 0$, which is precisely the cusp catastrophe model.

The above derivation from the fermionic imaginary-time path integral to cusp catastrophe incorporates the subtleties of Grassmann algebra, Nambu-Gorkov formalism, and particle-hole symmetry. The resulting cusp catastrophe model is universal, independent of the microscopic details of statistical quantum mechanics, which is exactly the power of structural stability and Thom's classification theorem.

## 5. Analysis of Superconducting Phase Transitions

Due to the strong interaction between electrons in high-temperature superconductors, the description of their electronic behavior inevitably involves the extremely complex many-body quantum physics, so the mechanism of high-temperature superconductivity is still not fully understood, especially the source of the attraction effect used for pairing between carriers is still the core of the current physics field one of the difficulties.

In our previous work [34], the thermodynamic quantum phase transition process has been investigated quantitatively by the structural-stability-based catastrophe theory. For a canonical ensemble composed of $N$ identical particles, the cusp catastrophe model is adopted to express the average free energy, as well as the canonical partition function and the specific heat capacity of the system.

Based on this thermodynamic quantum state equation, the adsorption potential for superconductors was presented [35]. Then we calculated the molar adsorption potentials for those typical superconductors, and found that it is almost proportional to the superconductivity temperature Tc.

Furthermore, the electron pairing mechanism by sharing the charge-regulated interfacial layer induced by the adsorption potential and exchanging the adsorption mode of high-temperature superconductors was revealed，and the effective interaction potential between electrons was derived in details [36]. In this work, we have verified that *d*-wave come from the anisotropy of interfacial adsorption forces, explained the pseudo-energy gap behavior, and obtained the coherence length expression. The quantitative predictions of the coherence length and superconducting gap close to the known results could verify our theoretical framework.

The structural stability theorem of Thom guarantees the universality of our results. Any quantum many-body system with control parameters $m \le 4$ undergoing a continuous phase transition must, in the critical regime, exhibit a bifurcation structure diffeomorphic to one of the seven elementary catastrophes. This universality explains why the catastrophe-theoretic framework successfully describes diverse systems ranging from liquid He-4 superfluidity to high-Tc cuprates.

Our analysis is restricted to the critical neighborhood where the linearization approximation holds. Far from Tc, non-uniform modes and higher-order gradient terms become significant, potentially requiring more complex catastrophe models (e.g., the butterfly catastrophe for tricritical behavior). The extension to dynamical phase transitions and non-equilibrium steady states remains an open challenge. Additionally, the precise microscopic derivation of the adsorption potential from first-principles electronic structure calculations would strengthen the predictive power of the theory.

## 6. Conclusion

In this work, we have established a rigorous mathematical framework connecting quantum many-body physics to catastrophe theory through the Lyapunov-Schmidt reduction procedure. Our results demonstrate that the cusp catastrophe structure is intrinsic to quantum many-body phase transitions, independent of microscopic details. This universality is protected by the structural stability theorem of Thom, which guarantees that any structurally stable gradient system with control parameters $k \leq 4$ must belong to one of the seven elementary catastrophe classes. This perspective unifies the Ginzburg-Landau-Wilson paradigm with Thom's structural stability theory, offering a non-perturbative, topologically protected framework for understanding superconductivity across materials classes—from conventional BCS superconductors to high-Tc cuprates, nickelates, and beyond.